\documentclass[twocolumn,showpacs,preprintnumbers]{revtex4} 
\usepackage{amssymb}
\usepackage[dvips]{graphicx}

\begin{document}

\title{Intrinsic tunnelling or wishful thinking?}

\author{V.N. Zavaritsky$^{1,2}$}
\address
{$^{1}$Department of Physics, Loughborough University, Loughborough, United Kingdom, 
$^{2}$Kapitza Institute for Physical Problems and General Physics Institute,  Moscow, Russia\\
}
\textbf{Intrinsic tunnelling or wishful thinking?}

Heating is known to distort I-V linearity in superconductors, cf. eg. \cite{gurevich}. However, heating issues in `intrinsic tunnelling' (IJT) were mostly misinterpreted or ignored until %recently, when 
direct experimental evidence of the heating origin of I-V nonlinearities in Bi2212 mesas was obtained \cite{gien}. Subsequently the origin of IJT and some principal %most important 
applications of the effect were addressed by \cite{prb} following the development of a practical tool for the distinction of intrinsic features from extrinsic ones and evidence of the heating origin of the key IJT findings including various IJT `gaps' \cite{comment2201}. Krasnov et al \cite{krasnov} ignored earlier experimental work by Refs.\cite{gurevich,gien,prb} and in undeclared polemics with Ref.\cite{prb} claimed their I-V are free of heating because the sample area dependence of these I-V appears to be at odds with the model by \cite{krasnov}. This model, however, is irrelevant %to the experiment by \cite{krasnov} as 
since it disregards convection and ignores the topmost {\it metal} electrode of the enhanced area and thermal conductivity, which is a principal heat-sinking channel in the mesas by \cite{krasnov}. As will be shown below, the %key findings %
`heat-free' I-Vs 
by \cite{krasnov} are most probably caused by heating. %\cite{prb}.% hence making irrelevant the conclusions by \cite{krasnov}. I will also address the remainder key points of this letter and will finally show that the conclusions by \cite{krasnov} are not beyond dispute. 

Heat load $P\equiv VI/A$ characterizes self-heating since the heat $VI$ dissipated inside the sample escapes through its surface area $A$. %Similarly to Refs.[3,4-6] in \cite{krasnov}, the characteristic loads $P\simeq4-8kW/cm^2$ that are required to achieve the IJT gaps in \cite{krasnov} exceed that of a domestic kettle by 2-3 orders, hence pointing to the heating origin of these findings. 
Similarly to Refs.[3,4-6] in \cite{krasnov}, the characteristic loads that are required to achieve the features in IJT spectra ascribed by \cite{krasnov} to different phenomena exceed that of a domestic kettle by 2-3 orders and are practically independent of $A$, hence %undoubtedly 
pointing to the heating origin of these features, see Fig.1(a). % and the applicability of the model by Refs.\cite{comment2201,prb}.

Mean overheating of a mesa-structure is described by Newton's law of cooling (1701) as 
\begin{equation}
T -T_B\propto P.
\end{equation}
%where $h$ is an area independent heat transfer coefficient, \cite{prb}. 
According to Eq.(1), monitoring I-V at a given bath temperature $T_B$ results in a sample temperature rise that entails  gap-like non-linearity if  $\partial R/\partial T$$<$0, Fig.1(b). Unlike the model by \cite{krasnov} the parameter-free Eq.(1) describes the %key IJT manifestations, including the 
measured IJT spectra {\it quantitatively} based on the $R(T)$ of {\it the same sample only} \cite{comment2201,prb}.  Here we extend the analysis by \cite{comment2201,prb} to verify whether the independence of the IJT gap from the sample area {\it guarantees} the absence of self-heating as claimed by \cite{krasnov}. Assuming the samples differ by the area {\it only}, we reconstructed the IJT spectra for the same range of $A$ using the model $R(T)$ by \cite{krasnov} reproduced in Fig.1(c). As seen from Fig.1(b), the thus constructed heating spectra reproduce all key findings by \cite{krasnov}. %Indeed, these spectra reveal the area independent IJT gap and some minor features, ascribed by \cite{krasnov} to phonons, and retain the same V-shape and slope below and above, irrespective of the mesa size. 
Indeed, irrespective of the mesa area, these spectra retain the same V-shape and slope below and above the IJT gap and even reveal some minor features ascribed by \cite{krasnov} to phonons. %Furthermore, to reach the IJT gap in Fig.1 the heating changes the conductance by a factor of three while the  differential conductance increases by one order in magnitude. 
Hence the principal claim by Krasnov et al is revealed to be invalid and the area independence of the shape of the IJT spectra by \cite{krasnov} shows that the latter indeed represent artefacts of self-heating. 

\begin{figure}
\begin{center}
\includegraphics[angle=-0,width=0.47\textwidth]{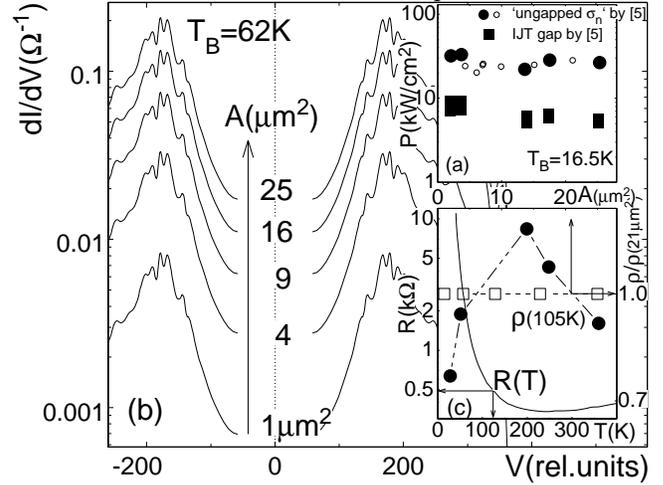}
\vskip -0.1mm
\caption{(a): Heat loads that are required to achieve the characteristic IJT features by \cite{krasnov}; (b): $\partial I/\partial V$ for mesas of different area made of the same medium, obtained from the model $R(T)$ of \cite{krasnov} (solid line in (c)) assuming Joule heating origin of I-V nonlinearities; (c) the squares and circles represent $\rho(A)$ of physically identical media and those by \cite{krasnov} respectively.  
}
\end{center}
\end{figure}

It is worth noting that Eg.(1) explains naturally the {\it actual} area `dependences' of IJT gaps by \cite{krasnov} that are {\it not flat}. %but reveal a noticeable scatter. 
%A straightforward 
Indeed, a direct estimate, $1/R=\partial I/\partial V|_{V\rightarrow0}$, shows that the samples by \cite{krasnov} reveal vastly different {\it resistivities} in stark contrast to what would be expected if these were physically identical, as is erroneously claimed by \cite{krasnov} and assumed in our calculations (see the symbols in Fig.1(c)). Hence Fig.2 by \cite{krasnov} becomes trivial as the IJT gaps of samples with different $\rho(T)$ should not necessarily match \cite{comment2201}.

To save space, I will only briefly point to some other key faults by \cite{krasnov} that are carried over from earlier works and were already addressed by \cite{comment2201,prb}. In particular, the IJT spectra at $T_B<T_c$ are described by the same Eq.(1) that explains naturally the abnormal sharpness of the gap-like feature at $T_B/T_c$$\sim$0.03, see \cite{prb} for details and also for the cause of I-V's back-bending that has little in common with heating above $T^*$. % but reflects $R(T)$ that is a material property. 
%Finally, the thermometry of the highly nonequilibrium state promoted by Krasnov et al is unreliable as it reproduces all the major faults of the predecessors, see \cite{comment2201,prb} for details.  
Finally, the nonequilibrium thermometry by \cite{krasnov} %Krasnov et al 
is unreliable as it ignores the heat load dependent thermal lag%between the sensor and the overheated mesa
, see \cite{comment2201,prb} for details.

In summary, our comprehensive analysis suggests the heating origin of key findings by \cite{krasnov}, hence rendering irrelevant their conclusions. 

V.N. Zavaritsky

Department of Physics, Loughborough University, UK; Kapitza Institute %for Physical Problems 
and General Physics Institute,  Moscow

\end{document}